\documentclass[superscriptaddress,12pt]{revtex4-2}
\usepackage[table]{xcolor}
\usepackage{rotating}
\usepackage{amsmath}
\usepackage{bm}
\usepackage{array}
\usepackage{graphicx}
\usepackage{dcolumn}
\usepackage{amssymb}
\usepackage[english]{babel}

\begin{document}

\title{The slow viscous flow around doubly-periodic arrays of infinite slender cylinders}

\author{Lyndon Koens\footnote{l.m.koens@hull.ac.uk}}
\affiliation{  Department of Mathematics, University of Hull, Hull HU6 7RX, United Kingdom}

\author{Rohan Vernekar}
\affiliation{Univ. Grenoble Alpes, CNRS, LRP, 38000 Grenoble, France}

\author{Timm Kr\"uger}
\affiliation{  School of Engineering, Institute for Multiscale Thermofluids, University of Edinburgh, Edinburgh EH9 3FB, United Kingdom}

\author{Maciej Lisicki}
\affiliation{ Institute of Theoretical Physics, Faculty of Physics, University of Warsaw, Warsaw, Pasteura 5, 02-093 Warsaw, Poland}

\author{David W. Inglis}
\affiliation{School of Engineering, Faculty of Science and Engineering, Macquarie University, Sydney, NSW 2109, Australia}

\begin{abstract}
The slow viscous flow through a doubly-periodic array of cylinders does not have an analytical solution. However, as a reduced model for the flow within fibrous porous media, this solution is important for many real-world systems. We asymptotically determine the flow around a doubly-periodic array of infinite slender cylinders, by placing doubly-periodic two-dimensional singularity solutions within the cylinder and expanding the no-slip condition on the cylinder's surface in powers of the cylinder radius. The asymptotic solution provides a closed-form estimate for the flow and forces as a function of the radius and the dimensions of the cell. The force is compared to results from lattice-Boltzmann simulations of low-Reynolds-number flows in the same geometry, and the accuracy of the no-slip condition on the surface of the cylinder, predicted by the asymptotic theory, is checked. Finally, the behaviour of the flow, flux, force and effective permeability of the cell is investigated as a function of the geometric parameters. The structure of the asymptotic permeability is consistent with other models for the flow parallel to an array of rods. These models could be used to help understand the flows within porous systems composed of fibres and systems involving periodic arrays such as deterministic lateral displacement.
\end{abstract}
\maketitle

\section{Introduction}
\label{sec:intro}

The slow viscous flow over multiple bodies is a notoriously complicated problem with various applications \cite{Kim2005}. For example, flowing colloids display discontinuous shear thickening \cite{Jamali2019, Wang2020}, and programmable self-assembling micromachines interact with each other through the flow to develop distinct phases and shapes \cite{Wang2022, Koens2019}. Similarly, filter-feeding organisms use the flow over a collection of microscopic fibres to capture food \cite{Nielsen2017, Lavrov2022, Blake1998}, and periodic arrays of posts in microfluidics can sort particles in a process called deterministic lateral displacement \cite{Inglis2006, Biagioni2020, Kim2017b, Jiang2016}.

These systems can be tricky to probe experimentally and are hard to model theoretically because slow viscous flows have long-ranged interactions \cite{Kim2005}. Even the relatively simplified geometries of a singularly or doubly-periodic array of infinite cylinders have no exact solutions. Yet, such model arrays have been studied since the late 1950s \cite{TAMADA1957} and have played an important role in understanding ordered or fibrous porous media \cite{Jackson1986}. Fibrous porous media, like wool, hair, collagen, and fibreglass, can exist at much lower packing fractions (below 1\%) than granular porous media (60-70\%) due to the large aspect ratios of the bodies.

The lack of exact closed solutions means the dynamics in singularly or doubly-periodic arrays of infinite cylinders are typically solved numerically \cite{Ayaz1999, Wang2001, Wang2002, Kirsh2006, BARTA2006a, Shou2015, Tran2022, Tran2022a} or argued from geometries with known solutions \cite{Maleki2017}. Through these methods, studies have explored the behaviour around cylinders in different periodic domains \cite{Wang2001, Kirsh2006, Maleki2017}, irregular domains \cite{Shou2015}, the influence of slip \cite{Wang2002}, cylinder porosity \cite{Kirsh2006}, interacting fluid domains \cite{Tran2022, Tran2022a}, and inertial effects \cite{Ayaz1999, TAMADA1957}.

Singularly periodic arrays have also been studied asymptotically. Such work provides a closed approximation to the solution that can be used when numerical approaches struggle, provide insight into how the geometry influences the behaviour and can be directly applied to new problems. Tamada and Fujikawa \cite{TAMADA1957} studied the drag on the cylinder when the periodic domain was much greater than the cylinder radius, while Keller \cite{Keller1964} considered the behaviour in the lubrication limit. Barta and Weihs  \cite{BARTA2006a} later used slender-body theory \cite{Koens2018, Keller1976a, Johnson1979} to improve the accuracy and investigate the effects of finite lengths and array size.

In contrast, relatively little asymptotic work has been done for doubly-periodic arrays. In the mid-1980s, Drummond and Tahir estimated the permeability of different periodic arrays through the matching of the flow outside an infinite cylinder and a collection of singularities \cite{Drummond1984}. Their tests suggest that the permeability, $k'$, for an array of cylinders typically has the form
\begin{equation}
\label{permmodel}
\frac{\mu k'}{R'^2} = \frac{1}{4 \phi}\left(- \log(\phi) + \alpha  - \beta \phi^2 + 2 \phi\right)
\end{equation}
where $R'$ is the radius of the cylinder, $\mu$ is the dynamic viscosity, $\phi$ is the packing fraction (the volume occupied by the cylinder divided by the total volume of the cell), and $\alpha$ and $\beta$ are constants that depend on the geometry of the cell \cite{Drummond1984, Jackson1986}. The values of $\alpha$ and $\beta$ need to be determined numerically for each geometry considered. Drummond and Tahir's models have been found to match several experimental results \cite{Jackson1986}. Wang \cite{Wang2001} later used the general solution to the flow in a periodic box and the solution for flow outside a cylinder to investigate the flow in the doubly-periodic arrays. They numerically enforced the boundary conditions to the flow at discrete locations on the edge of the domain to determine the unknown coefficients for each geometry they investigated. Consistent with Eq.~\eqref{permmodel}, they showed the leading logarithmic behaviour of the drag and the permeability for doubly-periodic arrays in the limit of small radius, but they also observed non-inertial vortices in front of and behind the cylinder when the radius increases. However, to the authors' knowledge, no closed-form asymptotic representation for the force and the flow in a doubly-periodic domain has been proposed to date.

This paper determines the asymptotic flow around and the force from a doubly-periodic array of cylinders, with dimensions $\ell'\times h'$, in the limit of small cylinder radius, $R'$. The solution is found by constructing a complex-variable two-dimensional singularity representation for the flow and satisfying the no-slip boundary condition on the cylinder surface up to $\mathcal{O}(R'^4/\ell'^4, R'^4/H'^4)$. The expansion provides a closed-form estimate for the flow and force in the cell in terms of the scaled radius of the cylinder and the aspect ratio of the two sides of the domain. The asymptotic force is compared to the force found from lattice-Boltzmann simulations for the same geometry at small Reynolds number, and the accuracy of the no-slip condition on the surface of the cylinder, as predicted by the asymptotic model, is investigated. The asymptotic solution allows us to create an analytical approximation for the mean velocity through the cell, the pressure drop across the cell, and the permeability of the system.

Section~\ref{sec:geo} introduces the doubly-periodic cylinder geometry considered. Sec.~\ref{sec:background} provides some background into the complex variable solutions of Stokes flow and introduces the doubly-periodic singularities used. These singularities are then employed in Sec.~\ref{sec:expand} to determine the asymptotic flow around the periodic cylinder system. The accuracy of these results is investigated in Sec.~\ref{sec:validation}. Finally, the flows predicted are discussed and the asymptotic permeability is determined in Sec.~\ref{sec:flow}, before concluding the paper in Sec.~\ref{sec:conclusion}.

\section{The infinite cylinder in a doubly-periodic cell}
\label{sec:geo}

\begin{figure}
\center
\includegraphics[width=0.95\textwidth]{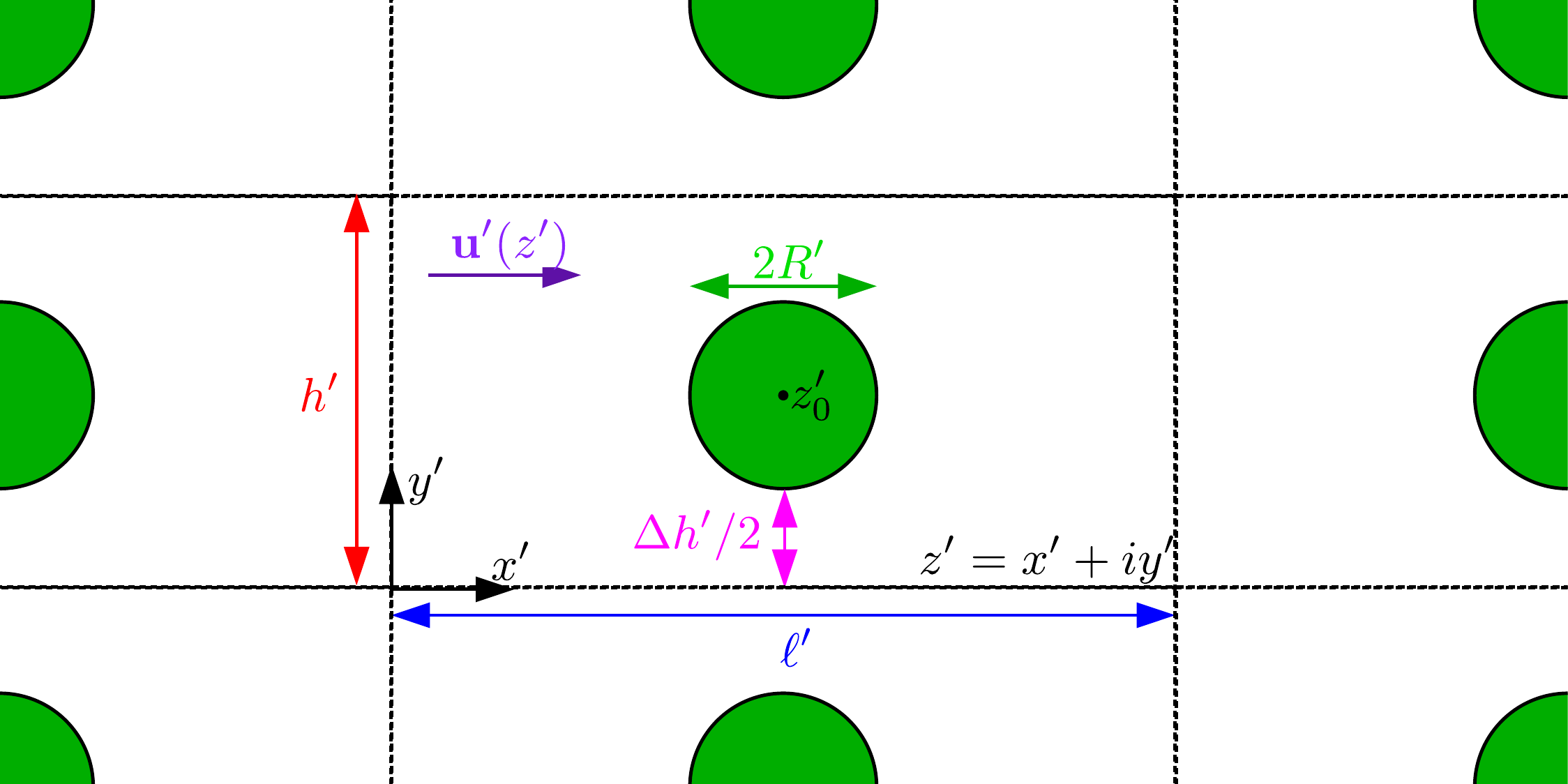}
\caption{A diagram of the doubly-periodic domain considered. The period along the $x'$-axis is $\ell'$, and the period along the $y'$-axis is $h'$. Any point in the domain on the complex plane is represented by $z' = x' + \text{i} y'$. The cylinder has a radius of $R'$ and is centred at $z_0'$. $\Delta h' = h'-2R'$ is the gap between two adjacent cylinders along the $y'$-axis. The flow is taken along the $x'$-axis without any loss of generality, because of the linearity of Stokes flow.}
\label{fig:diagram}
\end{figure}

This paper considers the slow viscous flow over an infinite slender cylinder of radius $R'$ in a doubly-periodic domain with periods $\ell'$ and $h'$, respectively (Fig.~\ref{fig:diagram}). The cylinder is assumed to be stationary, and the background flow will be taken along the $x$-axis. The dynamic viscosity of the fluid is $\mu$. In this geometry, the gap between two adjacent cylinders along the $y$-axis is given by $\Delta h' = h' - 2R'$. We scale all lengths by $\ell'$, velocities by the maximum velocity along the $x$-axis, $u_\text{m}'$, the force per unit length by $\mu u_\text{m}'$, the pressure by $\mu u_\text{m}'/\ell'$, and the permeability by $\ell'^2/\mu$. We note that, for the two-dimensional system, only the force per unit length is defined.

The flow within the unit cell is assumed to satisfy the incompressible Stokes equations
\begin{eqnarray}
 - \nabla p + \mu \nabla^2 \mathbf{u} &=& \mathbf{0},\\
  \nabla \cdot \mathbf{u} &=& 0
\end{eqnarray}
where $p$ is the pressure, and $\mathbf{u}$ is the velocity. The flow must also be periodic along the $x$- and $y$-axis, and the no-slip condition demands that the velocity should be zero on the surface of the cylinder.

As the geometry is two-dimensional, it is useful to solve for the flow on the complex plane (Fig.~\ref{fig:diagram}). Any point in the the scaled domain on the complex plane can be described by $z = x + \text{i} y$, where $x \in[0,1)$ and $y \in [0,h)$. The surface of the cylinder is given by
\begin{equation}
S = z_0 + R \text{e}^{\text{i} \theta}
\end{equation}
where $h = h'/\ell'$ is the aspect ratio of the cell, $R = R'/\ell'$ is the scaled cylinder radius, $z_0$ is the location of the centre of the cylinder, and $\theta$ is the polar angle from the real axis defined at the centre of the cylinder. The scaled gap between adjacent cylinders along the $y$-axis is $\Delta h = \Delta h'/\ell' = h-2R$. In the scaled coordinates, $R \in [0,\min(1/2,h/2))$ and $R/h \in [0,\min(1/(2h),1/2))$.

\section{Background in complex solutions to Stokes flow }
\label{sec:background} 

The incompressible Stokes equations are linear and time-independent. Solutions to these equations only depend on the instantaneous configuration of the system and can be constructed by the superposition of several flows. Solutions to the Stokes equations, with an appropriate set of boundary conditions, are also known to be unique. Even so, exact solutions to the Stokes equations are only known in relatively simple geometries. As such, many flows are approximated using numerical or asymptotic methods. These approaches often exploit the Green's function solution to the flow, called the Stokeslet. The Stokeslet represents the flow from a point force and can be used to construct solutions in two ways: the boundary integral method and the representation by fundamental singularities. On the one hand, the boundary integral method uses the properties of the Green's function to convert the partial differential equations into an integral equation over the boundaries of the domain. These equations can then be inverted numerically to determine the solution. On the other hand, the representation by fundamental singularities places the Stokeslet and its derivatives outside the flow domain such that the boundary conditions are satisfied \cite{Chwang2006}. In principle, such a distribution must exist \cite{Kim2005} and any suitable distribution must form the solution due to the uniqueness of the flow. The latter approach is often employed to find asymptotic solutions to common flow problems, such as the hydrodynamics of fibres \cite{Koens2018, Keller1976a, Johnson1979}.

In two dimensions, there is no solution to the Stokes equations for a point force in an unbounded domain \cite{Kim2005}. This so-called Stokes paradox is caused by the flow in two dimensions growing as $\log r$ with the distance $r$ from the point force. Solutions to Stokes flow in two dimensions, therefore, only exist in bounded domains or force-free unbounded domains. In such a domain, the flow can always be expressed in terms of a stream function, $\psi$, which is related to the flow velocity through
\begin{equation}
u = \frac{\partial \psi}{\partial y}, \quad v = -\frac{\partial \psi}{\partial x}
\end{equation}
where $\mathbf{u}=(u,v)$. The stream function satisfies the equation
\begin{equation}
\nabla^4 \psi = 0
\end{equation}
and has a general solution of the form 
\begin{equation}
\psi = \Im[ \bar{z} f(z) + g(z)]
\end{equation}
where $\Im[f(z)]$ returns the imaginary part of $f(z)$ and the overbar denotes the complex conjugate. $f(z)$ and $g(z)$ are analytic functions in the fluid region and are referred to as Goursat functions \cite{Langlois2014}. The flow generated from the above stream function is 
\begin{equation}
 u(z) - \text{i} v(z) = - \bar{f}(z) + \bar{z} \frac{df}{dz} + \frac{d g}{dz}.
\end{equation}
Though conformal maps do not preserve the boundary conditions for Stokes flow, this complex representation can be useful in determining the solutions to various problems \cite{Crowdy2019a, Crowdy2018, Luca2018, Crowdy2011}. However, no closed map exists for the flow around a cylinder in a doubly-periodic domain and, thus, the flow around an array of cylinders cannot currently be solved using conformal maps.

The two-dimensional flow from a point force per unit length in a doubly-periodic domain was first determined by Hasimoto \cite{Hasimoto1974}. Luca and Crowdy \cite{Crowdy2018} later revisited this problem to determine higher-order singularities and express it as a rapidly converging series. They showed that the flow from a two-dimensional point force per unit length of strength $-8 \pi F= -8 \pi (F_x+\text{i} F_y)$ located at $z_0 = x_0 + \text{i} y_0$ in a doubly-periodic cell with dimensions $x \in [0,1)$ and $y \in [0,h)$ \cite{Crowdy2018} can be written as
\begin{eqnarray}
 \label{stokeslet}
 u(z) - \text{i} v(z) = G_{S}(z-z_0,F) &=& - \bar{F} \ln |P(\zeta, \rho)|^2+ \Re[F] \ln|\zeta|^2 - F \ln|\zeta|^2 K(\zeta,\rho) \\ \notag
 && - F \ln (\rho^2) \rho \frac{\partial \ln P}{\partial \rho} - \frac{\Re[F]}{2 \ln \rho} (\ln|\zeta|^2)^2
\end{eqnarray}
where $\zeta = \exp[2 \pi \text{i} (z-z_0)]$, $\rho = \exp(-2 \pi h)$, $z = x + \text{i} y$ is the location in space and $F_x$ ($F_y$) represent the component $F$  in $x$ ($y$) direction. In Eq.~\eqref{stokeslet}, $P(\zeta,\rho)$ is the Schottky–Klein prime function associated with the annulus $\rho < |\zeta | < 1 $ \cite{Crowdy2018} and is given by
\begin{eqnarray}
 P(\zeta, \rho) &=& (1-\zeta) \prod_{k=1}^{\infty} (1- \rho^k \zeta) (1- \rho^k \zeta^{-1}) \\
 \label{rapid}
 &=& A(\rho) s(\zeta,\rho),
\end{eqnarray}
where
\begin{eqnarray}
 A(\rho) &=& \frac{\displaystyle \prod_{n =1}^{\infty} (1+\rho^n)^2}{\displaystyle \sum_{n=1}^{\infty} \rho^{n(n-1)/2}}, \\
 s(\zeta,\rho) &=&  \sum_{n=-\infty}^{\infty} (-1)^n \rho^{n(n-1)/2} \zeta^n.
\end{eqnarray}
Eq.~\eqref{rapid} is a rapidly convergent series representation that is useful to compute the flow \cite{Crowdy2018}. From the Schottky–Klein prime function $P(\zeta,\rho)$ we can define
\begin{eqnarray}
 K(\zeta,\rho) = \zeta \frac{\partial \ln P}{\partial \zeta} &=& \frac{\zeta s_{\zeta}}{s} ,\\
 \rho \frac{\partial \ln P}{\partial \rho} &=& \frac{\rho A_\rho}{A} +  \frac{\rho s_\rho}{s}
\end{eqnarray}
where the subscript denotes the derivative with respect to the given variable. Higher-order singularity solutions can be constructed from the Stokeslet by taking the appropriate derivatives. For example, the first derivative creates a force dipole, the second derivative creates a force quadrupole, etc. Similarly, the flow from a source dipole can also be constructed by taking the Laplacian of the Stokeslet flow \cite{Chwang2006}.

The symmetries of the doubly-periodic cell mean that, for our model, only singularities formed by an even number of derivatives of the Stokeslet will contribute to the flow. Section~\ref{sec:expand} will show that only the Stokeslet, force quadrupole, source dipole, and source octupole are needed to solve the flow up to $\mathcal{O}(R^4,(R/h)^4)$. These additional singularities are given by
\begin{eqnarray}  
 G_{Q}(z-z_0,Q) &=& - \overline{Q L(\zeta,\rho)}- 2 Q L - Q \ln|\zeta|^2 M(\zeta,\rho)- Q \ln (\rho^2) \rho \frac{\partial L}{\partial \rho}, \label{quad} \\
 G_{D}(z-z_0,D) &=&-D L(\zeta, \rho), \label{dipole} \\
 G_{O}(z-z_0,O) &=&  O N(\zeta,\rho), \label{octupole}
\end{eqnarray}
where $G_{Q}(z-z_0,Q)$ is the flow from a force quadrupole of the complex strength $Q = Q_x+\text{i}Q_y$ located at $z_0$, $G_{D}(z-z_0,D)$ is the flow from a source dipole of the complex strength $D=D_x+\text{i}D_y$ located at $z_0$, and $G_{O}(z-z_0,O)$ is the flow from a source octupole of the complex strength $O=O_x+\text{i}O_y$ located at $z_0$:
\begin{eqnarray}
 L(\zeta,\rho) = \zeta \frac{\partial K}{\partial \zeta} &=& \frac{\zeta [s (s_\zeta + \zeta s_{\zeta\zeta}) - \zeta s_\zeta^2]}{s^2} , \\
 M(\zeta,\rho) = \zeta \frac{\partial L}{\partial \zeta} &=& \frac{\zeta  (s- \zeta s_\zeta) [s (s_\zeta + 3\zeta s_{\zeta\zeta}) -2 \zeta s_\zeta^2] +\zeta^3 s^2 s_{\zeta\zeta\zeta}}{s^3} ,
 \\
 N(\zeta,\rho) = \zeta \frac{\partial M}{\partial \zeta} &=& \frac{\zeta  \left(12 \zeta ^2 s \left(s_\zeta+\zeta  s_{\zeta\zeta}\right) s_\zeta^2-\zeta  s^2 \left(3 \zeta ^2 s_{\zeta\zeta}^2+7 s_\zeta^2+2 \zeta  \left(9 s_{\zeta\zeta}+2 \zeta  s_{\zeta\zeta\zeta}\right) s_\zeta\right)\right)}{s^4}  \notag \\
 && +\frac{\zeta  \left( s^3 \left(\zeta  \left(\zeta ^2 s_{\zeta\zeta\zeta\zeta}+6 \zeta  s_{\zeta\zeta\zeta}+7 s_{\zeta\zeta}\right)+s_\zeta \right) -6 \zeta ^3 s_\zeta^4 \right)}{s^4},  \\
 \rho \frac{\partial L}{\partial \rho} &=& \frac{\zeta  \rho  \left(\left(s_{\zeta\rho}+\zeta  s_{\zeta\zeta\rho}\right) s^2-\left(2 \zeta  s_{\zeta} s_{\zeta\rho}+s_\rho \left(s_\zeta+\zeta  s_{\zeta\zeta}\right)\right) s+2 \zeta  s_\rho s_\zeta^2\right)}{s^3}. 
\end{eqnarray}
We have left these functions in terms of the derivatives of $s$ because they will be useful for the series expansion. A more compact representation for each function can be found in Ref.~\cite{Crowdy2018}.
 
\section{The asymptotic flow around slender cylinders} \label{sec:expand}

The flow around a cylinder in a doubly-periodic array can be determined in the slender limit ($R \ll 1$ and $R \ll h$) using the representation by fundamental singularities. In this method, singularity solutions are placed within the cylinder and their strength is chosen such that the boundary conditions are satisfied \cite{Chwang2006}. If such a solution can be found, it must be the solution, because of the uniqueness of Stokes flow.

Inspired by the seminal work of Chwang and Wu \cite{Chwang2006}, we seek a singularity representation in terms of a point force (Eq.~\eqref{stokeslet}), source dipole (Eq.~\eqref{dipole}), force quadrupole (Eq.~\eqref{quad}) and source octupole (Eq.~\eqref{octupole}) placed at the centre of the cylinder, $z_0$.
The flow from these singularities can be written as
\begin{equation}
 \label{flow}
 u(z) - \text{i} v(z) = U + G_{S}(z-z_0,F) + G_{D}(z-z_0,R^2 D) + G_{Q}(z-z_0,R^4 Q) + G_{O}(z-z_0,R^6 O) + \dots
\end{equation}
where $U$ is the unknown background flow speed in the absence of the cylinder, $-8 \pi F$ is the force per unit length on the fluid from the cylinder (strength of the Stokeslet), $R^2 D$ is the strength of the source dipole, $R^4 Q$ is the strength of the force quadrupole, and $R^6 O$ is the strength of the source octupole. The $R$ scaling in the strengths of the singularity is chosen to simplify the analysis, and the dots represent the higher singularities we are not considering. The proposed flow satisfies the doubly-periodic nature of the cell, so only the no-slip condition at the surface of the cylinder remains to be satisfied.

At the surface of the cylinder, the fluid velocity is zero and Eq.~\eqref{flow} becomes
\begin{equation}
 \label{flow2}
 0 = U + G_{S}(R \text{e}^{\text{i}\theta},F) + G_{D}(R \text{e}^{\text{i}\theta},R^2 D) + G_{Q}(R \text{e}^{\text{i}\theta},R^4 Q) + G_{O}(R \text{e}^{\text{i}\theta},R^6 O) + \dots
\end{equation}
Eq.~\eqref{flow2} needs to be solved to determine the unknown strengths of each of the singularities. Although a general solution cannot be obtained exactly, it is possible to find a solution in the limit of small scaled radius ($R \ll 1$ and $R \ll h$). In the small-$R$ limit, the singularities can be expressed as
\begin{eqnarray}
 G_{S}(R \text{e}^{\text{i}\theta}, F) &=& -F \left(G_{S}^{(0,0)} + R^2 G_{S}^{(2,0)} \right) + F \text{e}^{- 2 \text{i} \theta} \left(1 + R^2 G_{S}^{(2,-2)}\right)\notag \\
 &&+ F R^2 \text{e}^{2 \text{i} \theta} G_{S}^{(2,2)} + \mathcal{O}\left(R^4,\frac{R^4}{h^4}\right), \\
 G_{D}(R \text{e}^{\text{i}\theta},R^2 D) &=& -D \frac{\text{e}^{- 2 \text{i} \theta}}{4 \pi^2} -D \frac{R^2}{12} B(\rho)+ \mathcal{O}\left(R^4,\frac{R^4}{h^4}\right),\\
 G_{Q}(R \text{e}^{\text{i}\theta},R^4 Q) &=&- Q \frac{R^2 \text{e}^{2 \text{i} \theta}}{4 \pi^2} - Q \frac{R^2 \text{e}^{-4\text{i} \theta}}{2 \pi^2} + \mathcal{O}\left(R^4,\frac{R^4}{h^4}\right),\\
 G_{O}(R \text{e}^{\text{i}\theta},R^6 O) &=& - O \frac{3 R^2 \text{e}^{-4\text{i} \theta}}{8 \pi^4} + \mathcal{O}\left(R^4,\frac{R^4}{h^4}\right),
\end{eqnarray}
where
\begin{eqnarray}
 G_{S}^{(0,0)} &=& 1 +\ln\left[4 \pi^2 R^2 A^2(\rho) s_{\zeta}^2(1,\rho) \right] + 2 \rho \ln \rho \left[ \frac{A'(\rho)}{A(\rho)} + \frac{s_{\zeta,\rho}(1,\rho)}{s_{\zeta}(1,\rho)} \right], \\
 G_{S}^{(2,0)} &=& \frac{\pi^2}{3} \left[B(\rho) + \frac{12}{\ln \rho} \right],\\
 G_{S}^{(2,-2)} &=& \frac{\pi^2}{6} \left[B(\rho) + \frac{12}{\ln \rho} \right],\\
 G_{S}^{(2,2)} &=& \frac{\pi^2}{6} \left[3 B(\rho)+ \frac{12}{\ln \rho}+ 4 \rho \ln(\rho) \frac{d}{d\rho} \left(\frac{s_{\zeta,\zeta,\zeta}(1,\rho)}{s_{\zeta}(1,\rho)}\right) \right], \\
 B(\rho) &=& 1 -\frac{s_{\zeta,\zeta,\zeta,\zeta}(1,\rho)}{s_{\zeta}(1,\rho)}
\end{eqnarray}
and we have used the properties $s(1,\rho) =0$, $s_{\zeta,\zeta} (1,\rho) =0$, $s_{\zeta,\zeta,\zeta,\zeta} (1,\rho) +4s_{\zeta,\zeta,\zeta} (1,\rho)=0$. These properties are proven by separating the summations over $n$ into even and odd terms and noticing that the summation of the even terms is the negative of the summation of the odd terms when $\zeta=1$.
The above series representation expresses the singularities as powers of $\text{e}^{\text{i} \theta}$. Since different powers of $\text{e}^{\theta{i} \theta}$ are orthogonal, Eq.~\eqref{flow2} decomposes into four linear simultaneous equations for $F$, $D$, $Q$ and $O$. The solutions of these equations give us
\begin{eqnarray}
 F &= C_{F,U} U &=\frac{3 }{3G_{S}^{(0,0)} + R^2 \left[ \pi^2 B(\rho) +3 G_{S}^{(2,0)} \right]} U + \mathcal{O}\left(R^4,\frac{R^4}{h^4}\right) \label{F},\\
 D &= C_{D,F} F &= 4 \pi^2   \left(1 + R^2 G_{S}^{(2,-2)} \right) F + \mathcal{O}\left(R^4,\frac{R^4}{h^4}\right), \label{d}\\
 Q &= C_{Q,F} F &= 4 \pi^2  G_{S}^{(2,2)} F + \mathcal{O}(R^2), \label{q}\\
 O &= C_{O,Q} Q &= - \frac{4 \pi^2}{3} Q + \mathcal{O}(R^2) \label{o}
\end{eqnarray}
where $C_{F,U}$, $C_{D,F}$, $C_{Q,F}$, and $C_{O,Q}$ are the linearity coefficients that give the first subscript in terms of the second. The coefficient relating the Stokelet strength and the background velocity, $C_{F,U}$, is a scaled drag coefficient for the cylinder and behaves as $1/(\ln(R)+c)$, where $c$ is a constant, to leading order in $R$. The drag on slender rods often displays a similar $1/(\ln(R)+c)$ structure \cite{gray1955, Lighthill1987, Chwang2006} and is consistent with the limiting behaviour in previous studies of cylinders in periodic arrays \cite{Drummond1984, Wang2001}.

\begin{figure}[bt]
 \center
 \includegraphics[width=\textwidth]{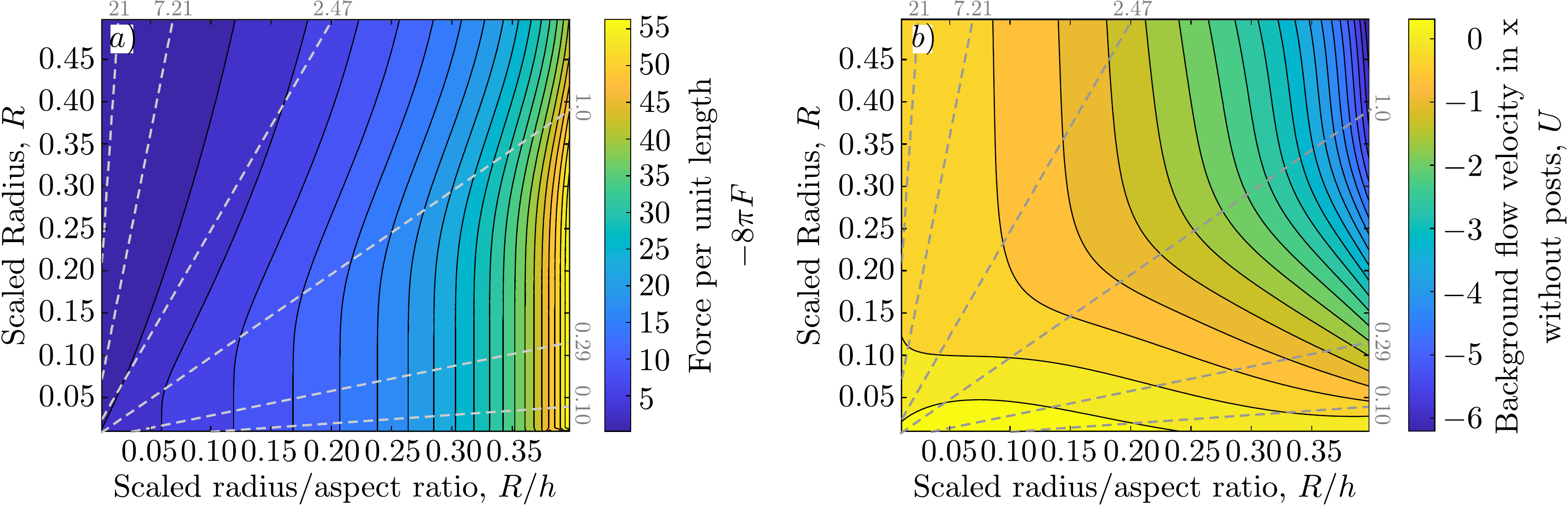}
 \caption{Contour plots of (a) force per unit length, $-8 \pi F$, and (b) the background flow in the absence of posts, $U$. Dashed grey lines are lines of constant cell aspect ratio, $h$.}
\label{fig:validation_contours}
\end{figure}

Finally, the strength of $U$ is found by setting the maximum velocity in the cell to 1. Conservation of mass dictates that the maximum velocity must lie on the edge of the domain directly above the post, at $z - z_0 = \text{i} h/2$. Hence $U$ satisfies
\begin{eqnarray}
 1 &=& U + G_{S}\left(\frac{\text{i} h}{2},C_{F,U}\right)U + G_{D}\left(\frac{\text{i} h}{2},R^2 C_{D,F} C_{F,U} \right) U + G_{Q}\left(\frac{\text{i} h}{2},R^4 C_{Q,F} C_{F,U}\right) U \notag \\
 && + G_{O} \left(\frac{\text{i} h}{2},R^6 C_{O,Q} C_{Q,F} C_{F,U}\right)U + \mathcal{O}\left(R^4,\frac{R^4}{h^4}\right)
\end{eqnarray}
or 
\begin{eqnarray}
 U &=& \left\{1 + \left[G_{S}\left(\frac{\text{i} h}{2},1\right) + G_{D}\left(\frac{\text{i} h}{2},C_{D,F} \right) + G_{Q}\left(\frac{\text{i} h}{2}, C_{Q,F} \right) \right.\right. \notag \\
 && \left. \left. + G_{O}\left(\frac{\text{i} h}{2},C_{O,Q} C_{Q,F} \right) \right] C_{F,U}\right\}^{-1} + \mathcal{O}\left(R^4,\frac{R^4}{h^4}\right) \label{U}
\end{eqnarray}
where we have used the fact that the singularity strengths are real. Eqs.~\eqref{F}, \eqref{d}, \eqref{q},\eqref{o} and \eqref{U} uniquely determine all the unknown coefficients in the proposed flow representation, Eq.~\eqref{flow}, to $\mathcal{O}(R^4,R^4/h^4)$. Hence, they complete the asymptotic model. The asymptotic model can be used to predict the force per unit length from the cylinder onto the fluid, $-8 \pi F$ (Fig.~\ref{fig:validation_contours}a), the background flow in the absence of the cylinders, $U$ (Fig.~\ref{fig:validation_contours}b), and the flow around the cylinder, Eq.~\eqref{flow}. We note that higher-order solutions can be constructed similarly but require higher-order singularities to satisfy the boundary conditions.

\section{Validation tests of the asymptotic model}
\label{sec:validation}

The asymptotic force per unit length on the fluid (Fig.~\ref{fig:validation_contours}a) decreases as the scaled radius, $R$, decreases and the cell aspect ratio, $h$, increases, as expected. As the cell aspect ratio decreases ($R/h$ increases), the cylinders become closer together, increasing the hydrodynamic interactions between them and the drag. Similarly, as the radius increases, the space between the cylinders decreases and the drag increases.

\begin{figure}
 \center
 \includegraphics[width=\textwidth]{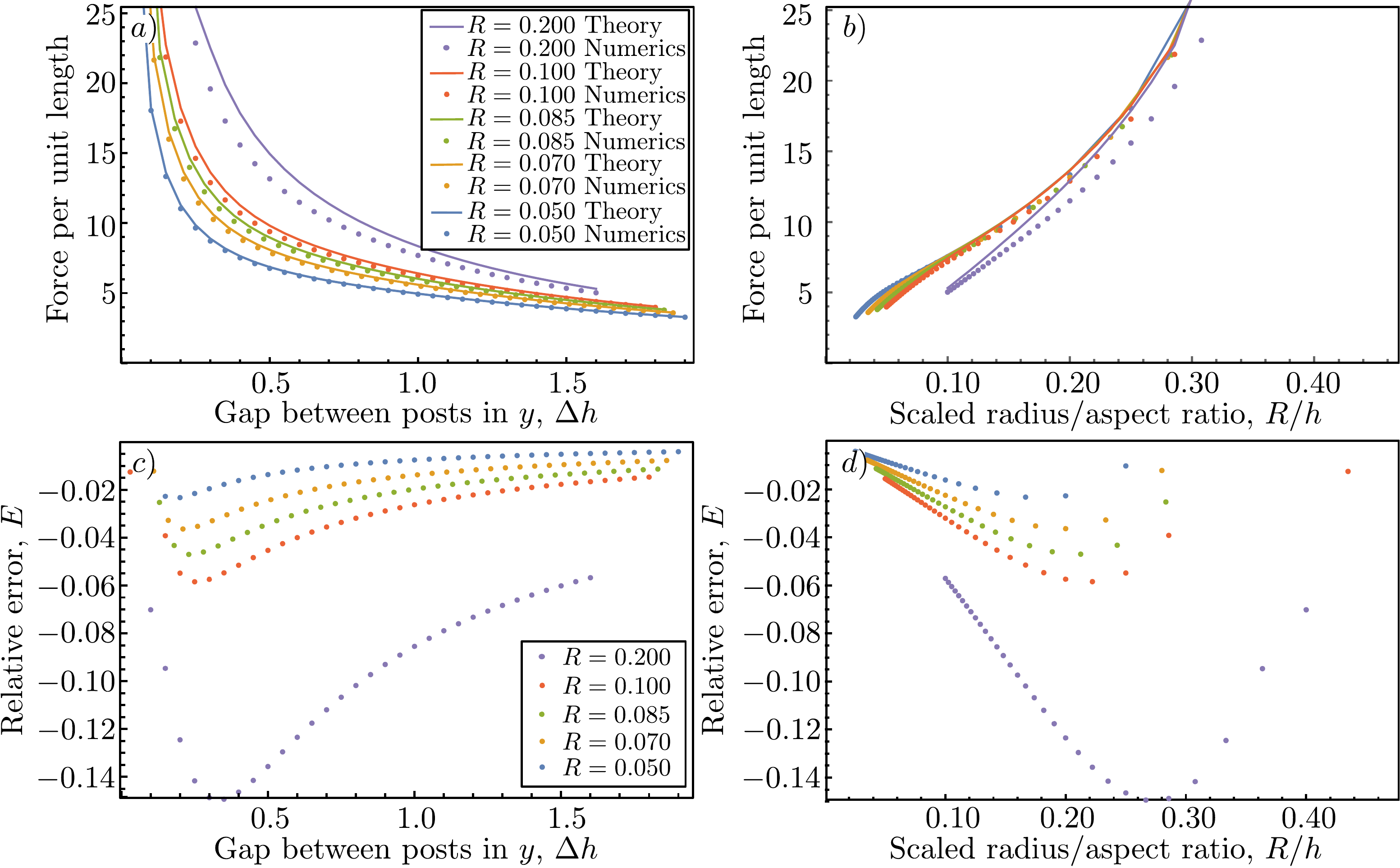}
 \caption{(a,b) Comparison of the asymptotic force per unit length,  $-8 \pi F$ (lines), and the numerical results of lattice-Boltzmann simulations (dots). (c,d) Relative error between the asymptotic force and the simulated force. The relative error is defined by Eq.~\eqref{rele}. (a) and (c) are plotted against the gap between adjacent posts along the $y$-axis, $\Delta h$. b) and c) are plotted against the scaled radius divided by the aspect ratio of the cell, $R/h$. In all panels, $R=0.2$ is purple, $R=0.1$ is red, $R=0.085$ is green, $R=0.07$ is yellow and $R=0.05$ is blue.}
 \label{fig:data}
\end{figure}

The accuracy of the asymptotic force per unit length, $-8 \pi F$ (Eq.~\eqref{F}), was quantified by comparing to the force per unit length determined from lattice-Boltzmann (LB) simulations for the flow past a single stationary post in a rectangular domain with periodic boundaries along both the $x$- and $y$-axis (Fig.~\ref{fig:data}) \cite{vernekar_anisotropic_2017}.
200 LB nodes along the $x$-axis were used for all simulated cases, while the number of nodes along the $y$-axis varied from 26 to 400.
The flow was driven by a constant body force along the $x$-axis, mimicking a constant pressure gradient.
The resolution varied from 20 to 80 LB nodes across the post diameter, with the stationary cylindrical boundary resolved using a second-order accurate boundary condition \cite{bouzidi_momentum_2001}.
We used the 9-velocity `compressible' LB equilibrium model in the limit of low Mach numbers ($Ma$) and Reynolds numbers ($Re$), with the two-relaxation time collision operator \cite{ginzburg_two-relaxation-time_2008}. The force on the post is computed with the momentum exchange algorithm \cite{ladd_numerical_1994, luo_numerics_2011}.
The resulting flow is practically incompressible and in the Stokes flow regime (since in all simulated cases, the LB Mach number $Ma < 1.9 \times 10^{-5}$ and the Reynolds number $Re < 1.7 \times 10^{-2}$, computed with the gap length, $\Delta h=h-2R$, and the maximum velocity magnitude), and each simulation has been run until the maximum velocity converges below $\le 0.01\%$ relative change per time step.
We did not see that the simulation results change significantly when both $Ma$ and $Re$ are varied by an order of magnitude.

The asymptotic force per unit length found in the asymptotic model,  $-8 \pi F$ (Eq.~\eqref{F}), and the LB simulation, $F_\text{LB}$, agree well for small scaled radii, $R$, and larger gaps between adjacent posts along the $y$-axis, $\Delta h$ (Fig.~\ref{fig:data}a). The differences are harder to distinguish when plotted against $R/h$ (Fig.~\ref{fig:data}b). The relative error, $E$, between the two results, defined as
\begin{eqnarray}
 \label{rele}
 E = \frac{F_\text{LB}-(-8 \pi F) }{F_{LB}},
\end{eqnarray} 
is shown in Fig.~\ref{fig:data}c,d. Similarly to direct plots of force, the relative error, $E$, increases with increasing scaled radius, $R$, and decreases with increasing cell aspect ratio, $h$. The results suggest that when $R<0.1$ and $R/h<0.1$ the asymptotic force displays less than 5\% error (Eq.~\eqref{rele}). These findings are consistent with the limits expected from the asymptotic expansion. We note that for small gap sizes, $\Delta h< 0.25$, the relative error starts to decrease again. These small gap sizes tend to correspond to $R/h>0.2$ and a rapid increase in $F$, indicating that it is well outside the region of validity for the model. The apparent improvement is therefore likely to be a coincidence.

Unlike the asymptotic force, the background velocity within the cell without the cylinder, $U$ (Fig.~\ref{fig:validation_contours}b), is less easy to understand. At small scaled radii, $R$, the background velocity is positive. As $R$ increases, the velocity decreases, ultimately changing its sign at some $R$. The rate of decrease increases with smaller $h$. It is important to recognise that the background velocity, $U$, is not the mean flow in the cell, but a mathematical construct necessary to scale the flow by the maximum velocity in the cell. It can never be observed numerically or experimentally. The mean flow in the cell comes from the combination of the background velocity and the net flows generated by the singularities (determined in Sec.~\ref{sec:flow}). Therefore, the background velocity changes its sign as the strengths of the singularities involved increase. As the incompressible Stokes equations are linear, an increase in the strengths of the singularities corresponds to an increase in the flux they generate. The background flow, therefore, changes its sign to reduce the mean flow and thereby keep the velocity everywhere within the cell below 1.

\begin{figure}[bt]
 \center
 \includegraphics[width=\textwidth]{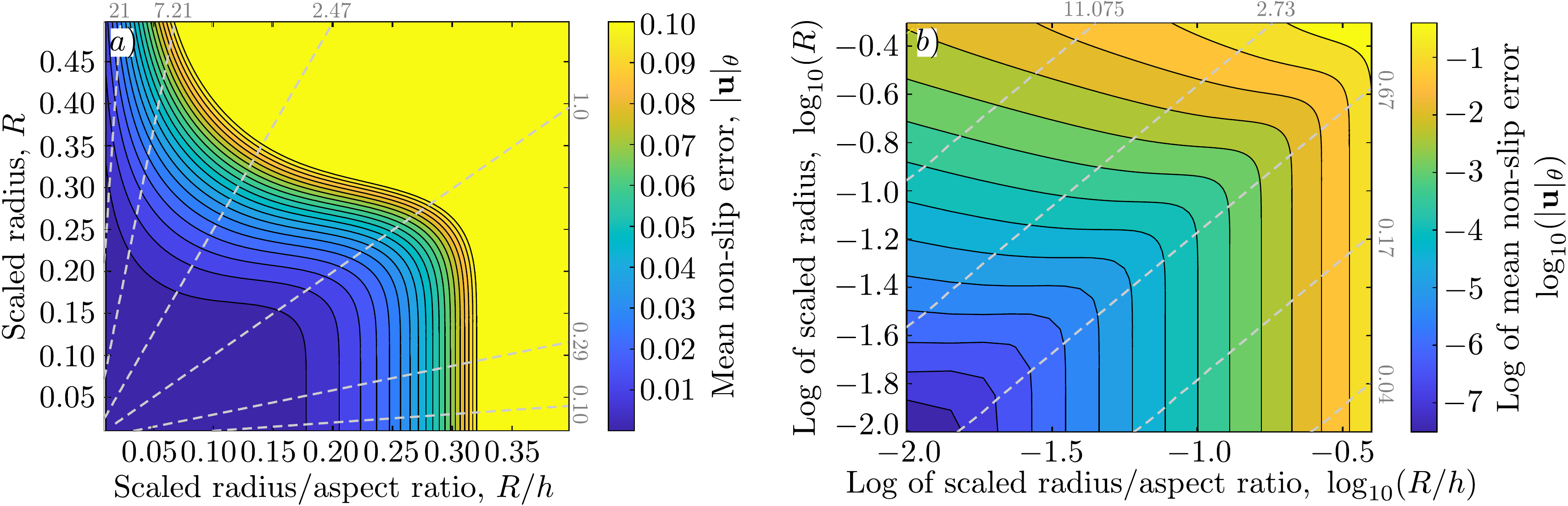}
 \caption{(a) Average error of the no-slip condition on the surface of the cylinder predicted by the asymptotic flow, Eq.~\eqref{flow}, as a function of the scaled radius, $R$, and the scaled radius divided by the cell aspect ratio, $R/h$. (b) Same as (a), but shown logarithmically. Dashed grey lines are lines of constant cell aspect ratio, $h$.}
 \label{fig:ENS}
\end{figure}

While the background velocity within the cell without the cylinder, $U$, is not useful for testing the accuracy of the predicted flow, the error on the asymptotic flow, Eq.~\eqref{flow}, can be estimated through the average error on the no-slip condition on the surface of the cylinder. The average error on the no-slip condition is defined as
\begin{equation}
 \label{div}
 |\mathbf{u}|_{\theta} = \frac{1}{2 \pi}\int_{0}^{2 \pi} \left|u(z_0+R \text{e}^{\text{i} \theta} ) - \text{i} v(z_0+R \text{e}^{\text{i} \theta} ) \right| \,d\theta
\end{equation}
and varies with the scaled radius, $R$, and the cell aspect ratio, $h$. If the asymptotic solution was exact, $|\mathbf{u}|_{\theta} = 0$, since Stokes flows are unique. Non-zero $|\mathbf{u}|_{\theta}$, therefore, indicates a difference between the unknown exact solution and the asymptotic prediction. The size of the difference must be proportional to $|\mathbf{u}|_{\theta}$ since Stokes flows are linear. Hence, small $|\mathbf{u}|_{\theta}$ corresponds to a small difference between the exact and the asymptotic flows. Contour plots of the average error on the no-slip condition are shown in Fig.~\ref{fig:ENS}. The average error on the no-slip condition increases with increasing scaled cylinder radius, $R$, and decreasing cell aspect ratio, $h$. On logarithmic axes (Fig.~\ref{fig:ENS}b), the increase appears to occur at a constant rate, suggesting power-law dependence on $R$ and $R/h$. The error of the asymptotic theory is expected to increase with $R^4$ and $(R/h)^4$ and so is consistent with the observed increase. A comparison of the relative error of the force per unit length, $E$ (Fig.~\ref{fig:data}c,d), and the average error on the no-slip condition suggests that when $|\mathbf{u}|_{\theta} < 0.01$ the error of the force per unit length is less than 6\%.
Similarly to the force, the 1\% error appears to occur around $R\sim 0.1$ and $R/h \sim 0.1$. For the remainder of the paper, all phase diagrams will consider $R<0.2$ and $R/h<0.2$ to focus on the region of validity of the asymptotic solution.

\section{The behaviour of the asymptotic flow}
\label{sec:flow}

\begin{figure}[bt]
 \center
 \includegraphics[width=\textwidth]{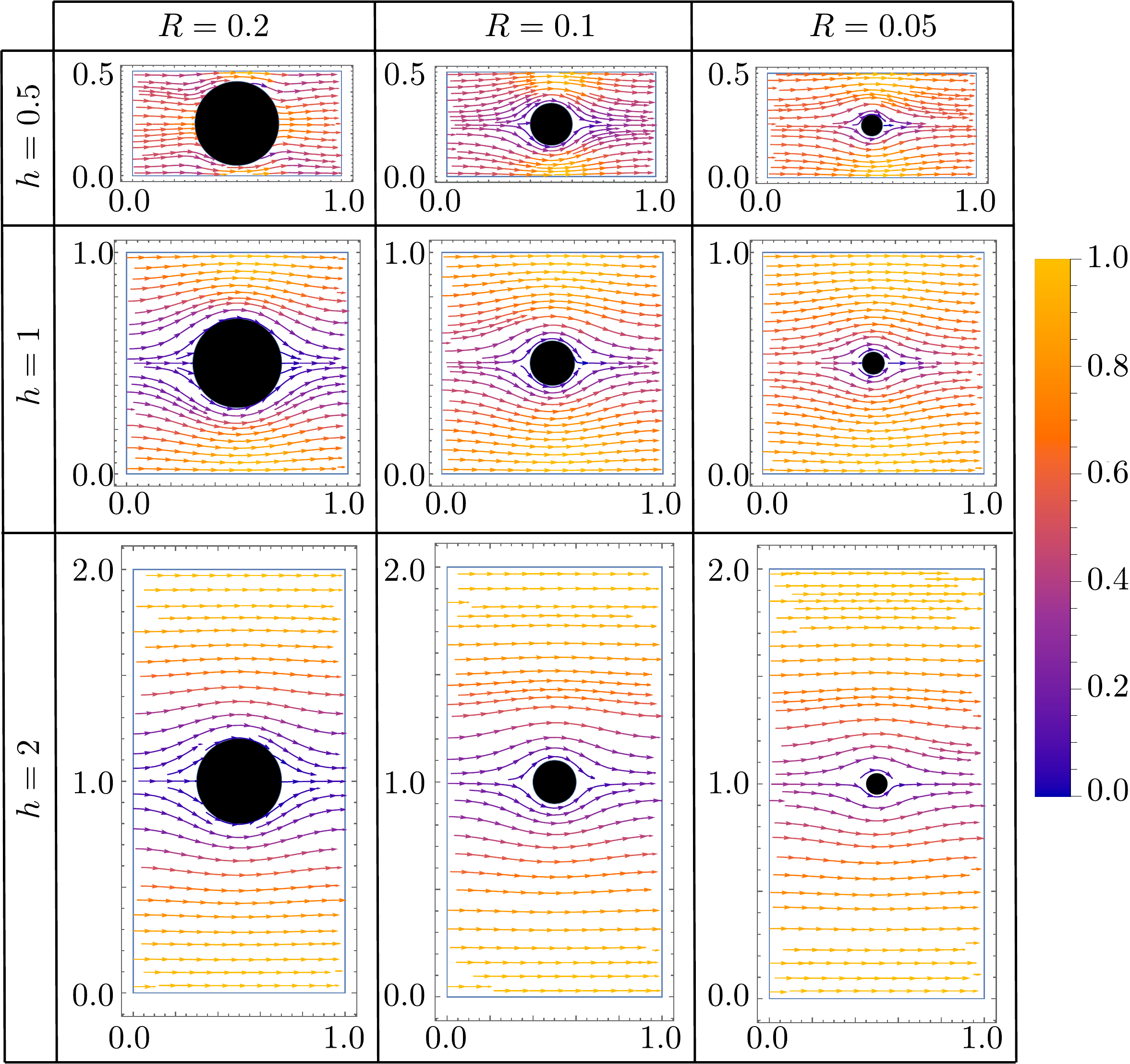}
 \caption{Streamlines for the flow predicted by the asymptotic solution (Sec.~\ref{sec:validation}) for scaled cylinder radii $R=0.2$, $0.1$, and $0.05$ and cell aspect ratios $h=2$, $1$, and $0.5$. The colour of the streamlines corresponds to the flow speed.}
 \label{fig:flows}
\end{figure}

The asymptotic model developed in Sec.~\ref{sec:expand} predicts the strengths of the singularities (Eqs.~\eqref{F}, \eqref{d}, \eqref{q}, and \eqref{o}), the background velocity in the absence of posts (Eq.~\eqref{U}), and the flow throughout the domain. The approximate flow at any point in the domain is given by Eq.~\eqref{flow} when the strengths of the singularities and the background velocity are substituted into the equation. The streamlines of the asymptotic flow are plotted in Fig.~\ref{fig:flows} for $z_0 = (1+ \text{i} h)/2$, and scaled radii of $R = 0.05$, $0.1$, and $0.2$ and cell aspect ratios of $h = 2$, $1$, and $0.5$. The scaled cylinder radius of $R=0.2$ lies beyond the validity limit of the model, established in the previous section, and is expected to show an error greater than 5\%, especially when the aspect ratio is $h=0.5$ ($R/h = 0.4$). Near each cylinder, the flow decreases. The size of the region with reduced velocity increases with scaled radius, $R$, and cell aspect ratio, $h$, due to the periodic interactions. The flows through cells with large aspect ratios also display a large portion of the fluid travelling at almost the maximum-velocity cell, while cells with smaller aspect ratios only reach the maximum velocity in a localised region directly above the cylinder. The localisation of the maximum velocity is caused by the conservation of mass which requires a faster velocity in the gap to squeeze the same amount of fluid through the cell. We note that the flow along the $x$-axis never reverses anywhere within the flow domain. Therefore, the asymptotic solution does not predict any closed vortices. Non-inertial vortices are typically found in Stokes flow at the leading and trailing edges of cylinders with large radius \cite{Wang2001}. The assumption that the cylinders are slender ($R \ll 1$ and $ R/h\ll 1$) in the expansion prevents the appearance of these vortices in the asymptotic flow.

The mean velocity in the periodic domain can be determined from the asymptotic flow by integrating the flow over $y =[0,h)$ for any $x$. The incompressible condition requires the flux through any plane normal to the $x$-axis to be the same. Hence the mean fluid velocity is given by
\begin{eqnarray}
 \langle u \rangle &=& \frac{1}{h} \int_{0}^{h} u(x+ \text{i} y) \, \text{d}y, \notag \\
 &=& U +\frac{1}{h}\int_{0}^{h}G_{S}(z-z_0,F) \, \text{d}y + \frac{1}{h}\int_{0}^{h} G_{D}(z-z_0,R^2 D) \, \text{d}y \notag \\
 && + \frac{1}{h}\int_{0}^{h} G_{Q}(z-z_0,R^4 Q) \, \text{d}y + \frac{1}{h} \int_{0}^{h}G_{O}(z-z_0,R^6 O) \, \text{d}y + \mathcal{O}\left(R^4,\frac{R^4}{h^4}\right)
\end{eqnarray}
where we chose $z_0 = (1 + \text{i} h)/2$. These integrals can be evaluated exactly using the properties of the Schottky–Klein prime function and its derivatives to find
\begin{equation}
 \label{meanv}
 \langle u \rangle = U + \langle G_{S} \rangle F + \langle G_{D} \rangle (R^2 D + R^4 Q) + \mathcal{O}\left(R^4,\frac{R^4}{h^4}\right)
\end{equation}
where
\begin{eqnarray}
 h \langle G_{S} \rangle &=& \frac{[\ln (\rho)]^2}{12 \pi}+ \frac{\ln \rho}{2 \pi} \ln \left( \frac{P^2 (-\sqrt{\rho},\rho)}{\sqrt{\rho}}\right) - \frac{\ln (\rho^2)}{ \pi} \sum_{k=1}^{\infty} k \ln\left(\frac{1+\rho^{k-1/2}}{1+\rho^{k+1/2}}\right) , \\
 h \langle G_{D} \rangle &=& \frac{ 2 K(-\sqrt{\rho},\rho)-1}{2 \pi}
\end{eqnarray}
are the fluid fluxes from a unit Stokeslet and source dipole, respectively. The flux from the source dipole is the same as the flux from a force quadrupole, while no flux is generated from the source octupole. The fluxes from each singularity depend only on the cell aspect ratio, $h$, through $\rho = \text{e}^{-2 \pi h}$. Hence, the dependence in the mean velocity on the scaled cylinder radius, $R$, comes from the background flow in the absence of the post, $U$, the force on the flow from the post, $F$, the source dipole strength, $R^2 D$, and the force quadrupole strength, $R^4 Q$. The background flow in the absence of the post, $U$, and the force on the flow from the post, $F$, depend on the scaled radius, $R$, logarithmically at leading order. Hence the mean velocity is expected to display a weak dependence on $R$ at leading order in the limit $R \ll 1$ and $R/h \ll 1$.

\begin{figure}
 \center
 \includegraphics[width=\textwidth]{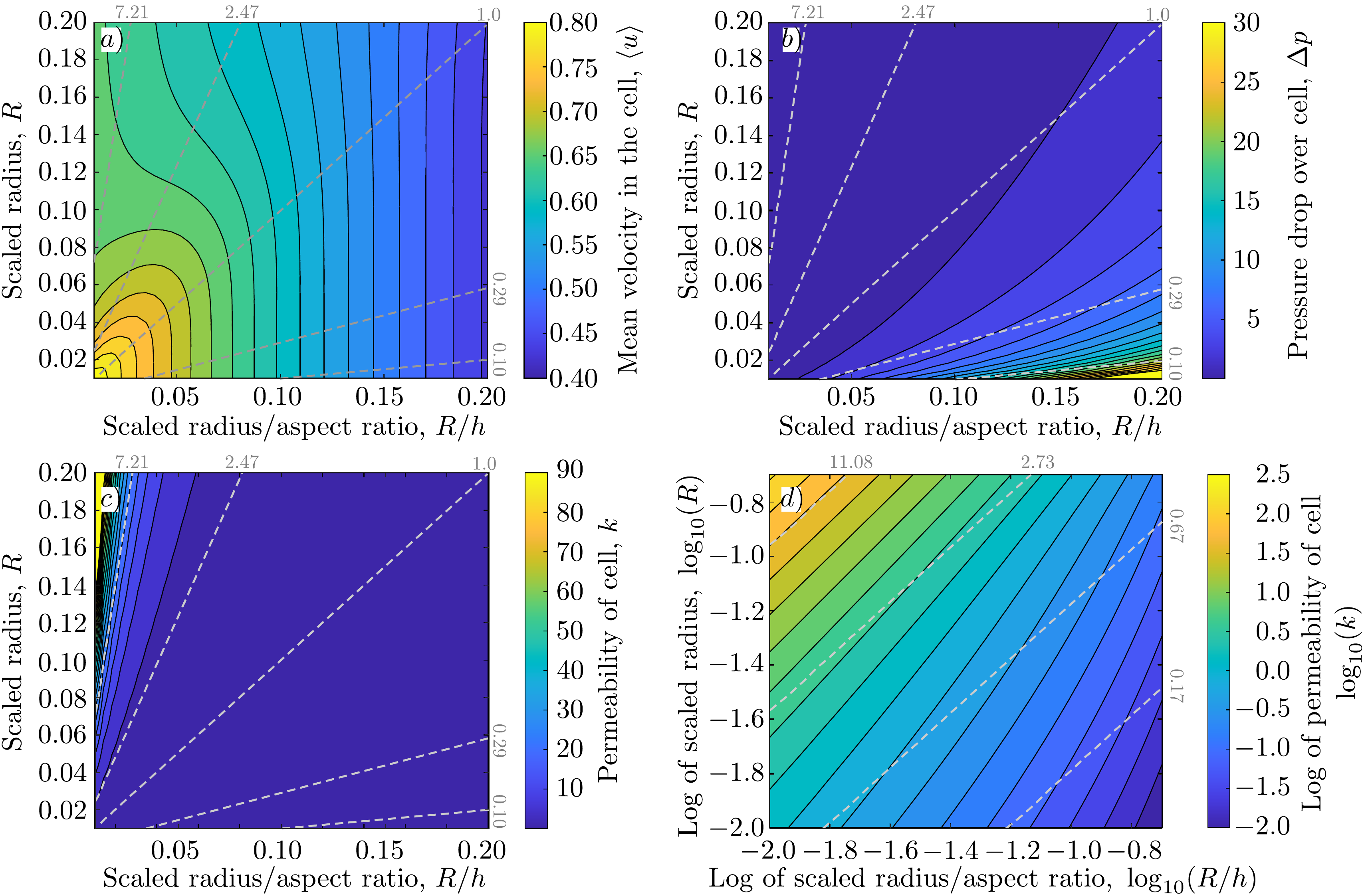}
 \caption{(a) Mean velocity in the cell, (b) pressure drop across the cell, and (c) permeability of the cell as a function of the scaled radius, $R$, and the scaled radius divided by the cell aspect ratio, $R/h$. (d) Same as (c), but shown logarithmically. Dashed grey lines are lines of constant cell aspect ratio, $h$.}
 \label{fig:mean}
\end{figure}

Although the leading mean velocity in $R$ is dominated by $\ln(R)$, the full asymptotic form, Eq.~\eqref{meanv}, shows a rich behaviour (Fig.~\ref{fig:mean}a). For small $R$ and $R/h$, the mean velocity is seen to increase as $R$ and $R/h$ decrease (increasing $h$ for constant $R$) because smaller $R$ and $R/h$ increase the separation between adjacent points, allowing more flow to travel near the maximum velocity of the cell. However, when $R/h>0.1$ or $R>0.1$ changes to the scaled radii tends to generate small changes in the mean velocity, $\langle u \rangle$, as the contours become vertical lines. It is unclear what the causes of this change are, but it may be related to issues in the asymptotic models at larger $R$ and $R/h$.

The pressure drop across the cell can also be determined from the asymptotic model (Fig.~\ref{fig:mean}b). The pressure drop in a doubly-periodic cell is given by $\Delta p =-4F/h$ where $F$ is the strength of the Stokeslet, Eq.~\eqref{F}. Similarly to the force, the pressure drop increases as the cell aspect ratio, $h$, decreases and the scaled radius, $R$, increases. The factor of $1/h$ in the definition, however, changes the behaviour with $R$, for fixed $R/h$, because $h$ must go to 0 to keep $R/h$ constant if $R \to 0$ and so the pressure starts to diverge in this limit. The physical understanding of the behaviour of the pressure drop is identical to that of the force.

The mean velocity in the cell and the pressure drop across the cell allow us to calculate the scaled permeability for the doubly-periodic array, $k$:
\begin{eqnarray}
 k = \frac{\langle u \rangle}{\Delta p} &=& -\frac{ h U + h \langle G_{S} \rangle F + h \langle G_{D} \rangle (R^2 D + R^4 Q)}{4 F} + \mathcal{O}\left(R^4,\frac{R^4}{h^4}\right) \notag \\
 &=& - \frac{h}{4 C_{F,U}} - \frac{h\langle G_{S} \rangle}{4}-\frac{h \langle G_{D} \rangle}{4}(R^2 C_{D,F} + R^4 C_{Q,F})+ \mathcal{O}\left(R^4,\frac{R^4}{h^4}\right) \notag \\
 &=& \frac{U}{\Delta p}- \frac{h \langle G_{S} \rangle}{4}-\frac{h \langle G_{D} \rangle}{4}(R^2 C_{D,F} + R^4 C_{Q,F})+ \mathcal{O}\left(R^4,\frac{R^4}{h^4}\right). \label{permeability}
\end{eqnarray}
The asymptotic permeability of the cell, Eq.~\eqref{permeability}, is dominated by the flux from a unit Stokeslet, which depends only on the aspect ratio $h$, and $U/\Delta p$, which depends on both the scaled radius $R$ and the aspect ratio $h$. Hence, at very small $R$ ($R\ll 1$ and $R/h \ll 1$), the permeability changes logarithmically with the radius, but it has a stronger dependence on the aspect ratio of the cell. Interestingly, if the permeability is written in terms of the packing fraction $\phi = \pi R^2/h$ instead of $R$, it can be expressed as
\begin{equation}
 \frac{\mu k'}{R'^2} = \frac{k}{R^2} = \frac{1}{4 \phi} \left( - \ln\phi +\alpha(h) + \beta(h) \phi^2  + \gamma(h) \phi\right) + \mathcal{O}\left(R^4,\frac{R^4}{h^4}\right)
\end{equation}
where
\begin{eqnarray}
 \alpha(h) &=& -1 -\langle G_{S} \rangle -\ln\left[4 \pi h A^2(\rho) s_{\zeta}^2(1,\rho) \right] - 2 \rho \ln \rho \left[ \frac{A'(\rho)}{A(\rho)} + \frac{s_{\zeta,\rho}(1,\rho)}{s_{\zeta}(1,\rho)} \right],\\
 \gamma(h) &=&  -\frac{h G_{S}^{(2,0)}}{\pi} - \frac{h \pi}{3}\left(B(\rho)+ 12 \langle G_{D} \rangle\right),\\
 \beta(h) &=& - 4 \langle G_{D} \rangle \left(G_{S}^{(2,-2)}+G_{S}^{(2,2)}\right).
\end{eqnarray}
The above form of the permeability is similar to the suggested form by Drummond and Tahir for regular arrays \cite{Drummond1984}, except all constants depend on the aspect ratio of the cell. The apparent difference is likely due to the different asymptotic approaches taken and the level of accuracy in the expansions. The asymptotic permeability, found above, therefore gives a closed form for these constants in doubly-periodic arrays.

The asymptotic permeability, Eq.~\eqref{permeability}, increases as the cell aspect ratio, $h$, increases and the scaled radius, $R$, decreases (Fig.~\ref{fig:mean}c,d). On a logarithmic scale, the lines of constant $\log_{10}(k)$ are almost straight lines (Fig.~\ref{fig:mean}d), suggesting that $\log_{10}(k) \propto a \log(R) - b\log(R/h)$ or $k \propto R^a (R/h)^{-b}$ where $a$ and $b$ are constants. The increase in asymptotic permeability with increasing $h$ and decreasing $R$ is due to the mean velocity in the cell, Eq.~\eqref{meanv}, having a much weaker dependence on $R$ and $h$ than the pressure drop across the cell, $\Delta p$ (Fig.~\ref{fig:mean}b), over the region of validity for the asymptotic model. Hence, when the pressure drop is small, the permeability becomes large.

\section{Conclusion}
\label{sec:conclusion}

This paper asymptotically determines the slow viscous flow around a doubly-periodic array of cylinders in the limit that the cylinder is slender. The slender condition means that the scaled radius, $R$, is much smaller than 1 and the cell aspect ratio, $h$. The asymptotic solution is constructed using the complex singularity solutions to two-dimensional Stokes flow and the representation by fundamental singularities. The results provide an analytical representation of the flow and the force per unit length as a function of the scaled radius of the cylinder, $R$, and the aspect ratio of the domain, $h$, and are accurate to $\mathcal{O}(R^4,R^4/h^4)$. The asymptotic force per unit length on the flow from the cylinder is compared to lattice-Boltzmann simulations for the same domain, and the accuracy of the asymptotically applied no-slip condition on the surface of the cylinder is investigated. The asymptotic results were then used to analyse the behaviour of the flow for varying scaled cylinder radius, $R$, and cell aspect ratio, $h$, and a closed representation of the permeability in the slender limit ($R \ll 1,h$) was developed. In the future, our analysis could be extended beyond $\mathcal{O}(R^4,R^4/h^4)$ through the addition of higher-order singularities. Our results could be of use in modelling the flows within porous systems composed of fibres and systems involving periodic arrays such as deterministic lateral displacement.

\section{Acknowledgments}

T.K.~received funding from the European Research Council (ERC) under the European Union’s Horizon 2020 research and innovation program (803553).
The work of ML was supported by the National Science Centre of Poland grant Sonata no. 2018/31/D/ST3/02408. 

For the purpose of open access, the authors have applied a Creative Commons Attribution (CC BY) licence to any Author Accepted Manuscript version arising from this submission.

\section{Data Management}
The MATLAB Implementation of asymptotic formulae used within the text and the data from the LB simulations plotted in fig~\ref{fig:data} are available on GitHub \cite{MPrograms}.

\bibliographystyle{ieeetr}
\bibliography{library_RV29Dec2022}

\end{document}